\documentclass[aps,prb,superscriptaddress,floatfix,twocolumn]{revtex4}
\usepackage{amssymb}
\usepackage{amsmath}
\usepackage{graphicx}

\begin{document}

\title{Electrical transport through a
  single-electron transistor strongly coupled to an oscillator}
\author{C. B. Doiron}
\affiliation{Department of Physics and Astronomy, University of Basel, 
CH-4056 Basel, Switzerland}
\author{W. Belzig}
\affiliation{University of Konstanz, Department of Physics, 
D-78457 Konstanz, Germany}
\author{C. Bruder}
\affiliation{Department of Physics and Astronomy, University of Basel, 
CH-4056 Basel, Switzerland}

\keywords{keywords}
\pacs{85.85.+j,73.23.-b,72.70.+m,73.23.Hk}

\begin{abstract}
We investigate electrical transport through a single-electron
transistor coupled to a nanomechanical oscillator.  Using a
combination of a master-equation approach and a numerical Monte Carlo
method, we calculate the average current and the current noise in the
strong-coupling regime, studying deviations from previously derived
analytic results valid in the limit of weak-coupling. After
generalizing the weak-coupling theory to enable the calculation of
higher cumulants of the current, we use our numerical approach to
study how the third cumulant is affected in the strong-coupling
regime. In this case, we find an interesting crossover between a
weak-coupling transport regime where the third cumulant heavily depends
on the frequency of the oscillator to one where it becomes practically
independent of this parameter.  Finally, we study the spectrum of the
transport noise and show that the two peaks found in the weak-coupling
limit merge on increasing the coupling strength. Our calculation of
the frequency-dependence of the noise also allows to describe how
transport-induced damping of the mechanical oscillations is affected
in the strong-coupling regime.
\end{abstract}

\date{\today}
\maketitle

\section{Introduction}
Nano-electromechanical devices, i.e., nanostructures in which electric
transport through a device is influenced by its mechanical degrees of
freedom and vice versa, have attracted a lot of interest
recently.\cite{roukes98,blencowe04} On the one hand, these devices are
promising for applications like sensors or ultra-sensitive mass
detectors. On the other hand, they have opened up new directions in
fundamental research, with projects to cool nanomechanical systems to
the quantum limit.\cite{roukes01,schwab04}

The nanomechanical properties of single-electron transistors (SETs)
are of particular interest in this context.  The central island of a
SET may be allowed to mechanically move between the two leads, such
that electrons can tunnel on the island if the island has approached
one lead and leave it again once it has mechanically moved to the
other lead. These ``shuttles'' have been investigated in great detail.
\cite{gorelik98,zwerger99,blick01,pistolesi04,isacsson04,jauho04,pistolesi06,fedorets04,fedorets05}
Another possibility to couple the electrical and mechanical properties
of the device is to design the SET such that its capacitive coupling
to the
gate depends on the displacement of a mechanical
oscillator. Thus, mechanical degrees of freedom of the system may
strongly influence the current-voltage characteristics, current noise,
and higher cumulants of the
current.\cite{armour04a,armour04b,imartin04,chtchelkatchev04a,jauho05,blanter06,fazio06}
It has been shown that the Coulomb blockade peaks are split for harmonic
oscillations and are broadened by thermal oscillations.  Knowledge of
the SET transport properties therefore allows one to determine 
the characteristics of the oscillator, such as
its amplitude and frequency. In such systems, electron tunneling
through the island also has an effect on the motion of the
oscillator. This back-action leads to fluctuations in the oscillator
position and to damping.\cite{clerk04} 

Practical implementations of oscillator-coupled SET transistors can be
realized by combining nanostructured silicon resonators with metallic
SETs.\cite{roukes96,knobel03a} Another possibility is to build SETs from suspended
carbon nanotubes that act as quantum dots.\cite{sapmaz03a} Quite
recently, mechanical oscillations
of the nanotube in such a device have been directly observed.\cite{mceuen}

In the following, we will consider a SET transistor coupled to a
classical harmonic oscillator. This system has already been studied
extensively.\cite{armour04a,armour04b} However, previous studies
investigated the regime where the coupling between the oscillator and
the SET is weak and the question what happens when the coupling is
increased is still of great theoretical 
interest,\cite{blencowe05a} even if this regime is 
not readily accessible in the current
generation of experiments. In this article, we will use a combination
of a master-equation approach and a numerical Monte Carlo procedure to
calculate the electrical current and its second and third cumulants
and study how they are modified by coupling to the oscillator, in the
regime where the coupling is strong. We will also study the frequency
dependence of the transport noise.

The paper is organized as follows: in Section \ref{SET-oscillator}, we
discuss the system and the model whose strong-coupling limit will be
studied in the subsequent sections.  The model and the master-equation
approach that we use follow closely Ref.~\onlinecite{armour04a}.
This section also introduces the important dimensionless coupling
parameter $\kappa$ that is the ratio of the typical mechanical energy
scale and the source-drain voltage. In the next section,
Sec.~\ref{dynamics_oscillator}, we calculate the probability
distributions of the position of the oscillator if the SET is in state
$N$ or $N+1$ using a numerical Monte Carlo procedure. The Gaussian
form predicted by the weak-coupling approach is modified dramatically
in the strong-coupling regime.  In Section \ref{average_current}, we
calculate the average current through the device at the
charge-degeneracy point of the SET at which the current is maximal and
discuss the deviations from the weak-coupling results. Finally, in
Sections \ref{noise_higher_moments} and
\ref{frequency-dependent_noise} we extend our studies to the current
noise and the third cumulant of the current.

\section{Coupled SET-oscillator system description}
\label{SET-oscillator}

To describe the coupled SET-nanomechanical oscillator system, we use
the formalism introduced in Ref.~\onlinecite{armour04a}.

\begin{figure}
\begin{center}
\includegraphics[scale=2.0]{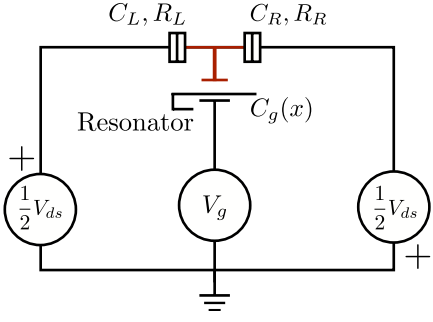}
\caption{Circuit diagram of the system studied. 
The gate capacitance of the SET depends on
the displacement of a mechanical oscillator, leading to a coupling
of the electrical transport through the device and the mechanical
motion of the oscillator.
}
\label{schematic}
\end{center}
\end{figure}

The system that we consider is shown in Fig.~\ref{schematic} in a
schematic way and consists of two symmetric tunnel junctions, each
with resistance $R$ and capacitance $C$, connected in
series. Transport through the SET is described using the orthodox
model, where only two charge states are considered and where the
current arises only from sequential
tunneling.\cite{averin91a,beenakker91a} In this case, transport is
governed by four tunneling rates $\Gamma _i ^\sigma$ where $i=R,L$ is the
lead index and $\sigma = +,-$ indicates the direction of the tunneling. In
this work, we adopt the convention that the forward $(+)$ direction, given
by the polarity of the bias voltage, is from the right to the left
lead. The tunneling rates can be calculated using Fermi's golden rule
and are a function of the difference in free energy $\Delta E$ of the
system before and after
a tunneling event
\begin{equation}
\Gamma _i ^\sigma = \frac{1}{Re^2} \Delta E _i ^\sigma \, 
f(\Delta E _i ^\sigma)\; ,
\end{equation}
where $f(x) = (1-e^{-x/k_B T})^{-1}$, with $T$ the electronic
temperature. The energy differences $\Delta E _i ^\sigma$ are given by
\begin{align}\label{delta_e}
\begin{split}
\Delta E ^+ _L &= -\Delta E ^- _L = e V_{ds} \Big(\frac{1}{2} + (2N -
2N_g +1) \frac{E_c}{eV_{ds}} \Big) \\ 
\Delta E ^+ _R &= -\Delta E ^- _R
= e V_{ds} \Big( \frac{1}{2} -(2N - 2N_g +1) \frac{E_c}{eV_{ds}}
\Big)\; ,
\end{split}
\end{align}
where $V_{ds}$ is the applied drain-source voltage, $E_c=e^2/(2C+C_g)$ is the
charging energy of the island and $N_g = C_g V_g / e$ is the optimal
number of charges on the island. Knowing the different rates, the average current $I$ 
flowing through the SET can be calculated using
\begin{equation}\label{current normal set}
I/e =  P_{N+1} \Gamma_L^+ - P_N \Gamma_L^- =  P_{N} \Gamma_R^+ -
P_{N+1} \Gamma_R^- \; ,
\end{equation}
where $P_{N(N+1)}$ is the probability to find the island in charge
state $N(N+1)$ in the stationary limit.

Our model of the SET remains valid as long as its charging energy
$E_c$ is large compared to the electronic thermal energy $k_B T_e$ and
the source-drain bias $ e V_{ds}$. We will neglect 
second-order tunneling processes (cotunneling).

In this work, the nanomechanical oscillator is modeled as a single,
classical, harmonic oscillator of frequency $\omega_0$. Introducing a time
scale $\tau_t = R e / V_{ds}$ 
which has the physical meaning of an average time between tunneling
events, we can use the dimensionless parameter 
\begin{equation}
\epsilon = \omega_0 \tau_t=\omega_0\frac{Re}{V_{ds}}
\label{epsilon}
\end{equation}
to compare the typical electrical and mechanical timescales.

A particular state of the oscillator is then represented by a
position $x$ and velocity $u$. We choose $x=0$ to be the equilibrium
point of the oscillator when $N$ charges are on the SET. When the
charge state of the island is changed, for example, from $N$ to $N+1$,
the change in the electrostatic forces between the oscillator (kept at
constant potential $V_g$) and the SET effectively shifts the
equilibrium position of the resonator. The distance between the
equilibrium positions when $N$ and $N+1$ charges are on the island
defines a natural length-scale $x_0$ of the problem, $x_0 = -2 E_c N_g
/ (m \omega_0^2d)$. Here, $d$ is the distance separating the
oscillator's equilibrium position and the SET island, such that the
gate capacitance depends on $x$ like $C_g(x)\sim (d+x)^{-1}\sim 1-x/d$.
From now on, we will also use
dimensionless rates, i.e., all the rates will be given in units of
$\tau_t ^{-1}$.

Coupling a SET and a nanomechanical oscillator system is readily done
by using the oscillator itself as the SET's gate. In this
configuration, the capacitive coupling between the oscillator and the
SET depends on the distance between them and by extension on the
oscillator's position, effectively allowing one to monitor the
dynamics of the oscillator via the SET. As long as the amplitude of
the oscillations around its equilibrium position is small compared to
the distance $d$ separating the oscillator and the SET island, the
gate capacitance $C_g(x)$ can be treated as linear in $x$. As a
consequence, we obtain position-dependent dimensionless tunneling
rates of the form
\begin{align} \label{eq rate} 
\begin{split}
\Gamma_L ^{+(-)} (x) &= (-) \big[ \Delta_L - \kappa \frac{x}{x_0}
\big] f \Big( (-) \left[ \Delta_L - \kappa \frac{x}{x_0} \right]
eV_{ds} \Big)
\\ \Gamma_R ^{+(-)} (x) &= (-) \big[ \Delta_R +
\kappa \frac{x}{x_0} \big] f \Big( (-) \left[ \Delta_R + \kappa
\frac{x}{x_0} \right]eV_{ds} \Big)\; ,
\end{split}
\end{align}
where the coefficients\cite{footnote1} 
\begin{align}\label{rates}
\begin{split}
\Delta_L &= \frac{1}{2} + (2N - 2N_g +1) \frac{E_c}{e V_{ds}} -\kappa N
\\
\Delta_R &= \frac{1}{2} -(2N - 2N_g +1) \frac{E_c}{e V_{ds}}+ \kappa N
\end{split}
\end{align}
are the position-independent part of the full
dimensionless rate $\Gamma_i ^\sigma(x)$ that fulfill
$\Delta_L+\Delta_R=1$, and 
\begin{equation}
\kappa = m \omega_0^2 x_0^2 / (eV_{ds}) 
\label{kappa}
\end{equation}
is a dimensionless coupling parameter that will play an important role
in the following.  Note that $\Delta_L$, $\Delta_R$ can become
negative in the strong-coupling limit. The average dimensionless
current in the presence of position-dependent rates can be calculated
as an average of the different rates weighted by the probability to
find the oscillator at a position $x$:
\begin{align}
I  &= \int^\infty _{- \infty} dx \,\big( P_{N+1} (x) 
\Gamma_L ^+ (x) - P_{N} (x) \Gamma _L ^- (x) \big) \nonumber\\ 
&= \int^\infty _{- \infty}
dx \, \big( P_{N} (x)  \Gamma_R ^+ (x) - P_{N+1} (x)  \Gamma _R ^- (x) \big)\; ,
\end{align}
with $P_{N(N+1)} (x)$ the probability to find the oscillator at
position $x$ while the island charge state is $N(N+1)$.

In the zero-temperature limit, the Fermi functions in 
Eqs.~(\ref{eq rate}) are in fact Heaviside step functions that
determine the possible transport direction as a function of the
position of the oscillator. Indeed, at zero temperature, $x^L =
\Delta_L x_0 / \kappa $ and $x^R = -\Delta_R x_0 / \kappa$ define
points where the current direction at lead $L$ and $R$ 
changes sign. For $x^R <x$ current in the right junction can only be
directed towards the island while in the opposite case only charge
transfer from the island to the right lead is possible. Equivalently,
transfer through the left junction is allowed from the island to the
lead if $x< x^L$ and from the lead to the island otherwise. It is
interesting to note that transport can be blocked altogether via this
mechanism. For example, if $N+1$ electrons are on the island and the
oscillator is in position $x>x^L$, transport of the extra charge from
the island to any lead is effectively forbidden, our choice of bias
direction imposing $x^R < x^L$.

The canonical way of dealing with an SET in the sequential tunneling
regime is to introduce a master equation for the different charge
states of the island. If the oscillator is coupled to a
nanomechanical oscillator, such a simple master equation cannot 
be written, since the tunneling rates depend on the
stochastic evolution of the oscillator. 
Following Ref.~\onlinecite{armour04a} we can introduce the
probability distributions $P_N(x,u;t)$ and $P_{N+1} (x,u;t)$ to find
at a time $t$, the oscillator at position $x,u$ in phase space and the
SET in charge state $N$ and $N+1$ respectively and derive a master
equation for these new objects:
\begin{subequations}
\label{master equation}
\begin{align}
\frac{\partial }{\partial t} P_N (x,u;t)&= \omega_0^2 x
\frac{\partial} {\partial u}P_N(x,u;t)- u \frac{\partial }{\partial x}
P_N(x,u;t) \nonumber\\ 
&+\left[ \Gamma _L^+ (x) + \Gamma_R ^- (x) \right]
P_{N+1}(x,u;t)\nonumber\\ 
&- \left[ \Gamma _R^+ (x) + \Gamma_L ^- (x) \right]
P_{N}(x,u;t)\; ,
\label{master equation pn}
\end{align}
\begin{align}
\frac{\partial }{\partial t} P_{N+1} (x,u;t)&= \omega_0^2 (x-x_0)
\frac{\partial} {\partial u}P_{N+1}(x,u;t)
\nonumber\\ 
&- u \frac{\partial }{\partial x} P_{N+1}(x,u;t) \nonumber\\ 
&-\left[ \Gamma _L^+ (x) + \Gamma_R ^-
(x) \right] P_{N+1}(x,u;t)\nonumber\\ 
&+\left[ \Gamma _R^+ (x) + \Gamma_L^- (x) \right] P_{N}(x,u;t)\; .
\label{master equation pnp1}
\end{align}
\end{subequations}
As pointed out in Ref.~\onlinecite{armour04a}, when the coupling
between the oscillator and the SET is weak ($\kappa \ll 1$) and when
the gate voltage $V_g$ is such that the system is tuned far from the
Coulomb-blockade region, one can make the approximation that $x^L \to
\infty$ and $x^R \to -\infty$ and then write the tunneling
rates as
\begin{align}\label{approximations}
\begin{split}
\Gamma _L ^+ (x)= \Delta_L - \kappa \frac{x}{x_0} &\; , \qquad
\Gamma_L^- (x) = 0\; ,\\
\Gamma_R^+ (x) = \Delta_R + \kappa \frac{x}{x_0} &\; , \qquad
\Gamma_R^- (x)= 0\; .
\end{split}
\end{align}
This weak-coupling form of Eq.~(\ref{eq rate}) effectively corresponds
to neglecting any back-currents and the possibility of
position-induced current blockade. However, the master equation is
then simple enough to allow analytical solutions.

In this work, we will not study the effect of extrinsic damping (i.e.,
a finite quality factor of the oscillator) and of finite temperatures,
since they were discussed extensively in
Refs.~\onlinecite{armour04a,armour04b}.

\section{Dynamics of the oscillator in the strong-coupling regime}
\label{dynamics_oscillator}

In the weak-coupling limit $\kappa \ll 1$, it was
found\cite{armour04a} that the interaction between the SET and the
oscillator introduces an intrinsic damping mechanism. The damping,
characterized by a decay rate $\gamma = \kappa\epsilon^2$ (in units
of $\tau_t^{-1}$) leads to a steady-state solution for the probability
distributions $P_{N(N+1)} (x,u)$. In particular, the probability
distributions $P_{N(N+1)} (x) = \int du \, P_{N(N+1)} (x,u)$, from
which one can calculate the average current, have been shown to be
well approximated by Gaussians centered at $x=0$ and $x=x_0$ for $P_N$
and $P_{N+1}$, respectively.

One of the main goals of this work is to study deviations from the
weak-coupling behavior. Without the simplifications possible for
$\kappa \ll 1$ leading to Eq.~(\ref{approximations}), the stationary
probability distributions $P_{N(N+1)} (x,u)$ can no longer be
investigated analytically and numerical methods must be used. In this
work, we used a Monte Carlo approach to simulate the stochastic nature
of the SET-nanomechanical oscillator system in the parameter regime
where the typical mechanical energy $m \omega_0^2 x_0^2$ is comparable
to the bias energy $ e V_{ds}$.  Details of our implementation of the
Monte Carlo method are given in Appendix \ref{MCmethod}.

We study the probability distribution of the oscillator in the
charge-degenerate case, where $\langle P_N \rangle = \langle P_{N+1}
\rangle$, 
where $\langle P_N \rangle =\int dx P_N(x)$.
At this point the current flowing through the SET is
maximal. In the presence of the oscillator, charge degeneracy is
reached when $\Delta_L = 1/2+\kappa/2$.
This relation, exact in the weak-coupling limit, has been empirically  
verified over the whole range of $\kappa$ studied. In the 
weak-coupling limit, this relation can be shown using $\langle P_{N+1}  
\rangle = \Delta_R $ familiar from 2-state SETs.   In our case, at  
the degeneracy point symmetry considerations impose $\langle x  
\rangle = x_0/2$ and the position dependence of the rate $\Gamma_R^+  
(x)$ must be taken into account, such that $\langle P_{N+1}  
\rangle =\frac{1}{2} = \Delta_R + \kappa \langle x / x_0\rangle$. This  
effectively corresponds to $ \langle P_{N+1} \rangle =\Delta_L -  
\kappa/2$.

\begin{figure}
\begin{center}
\includegraphics[scale=1.0]{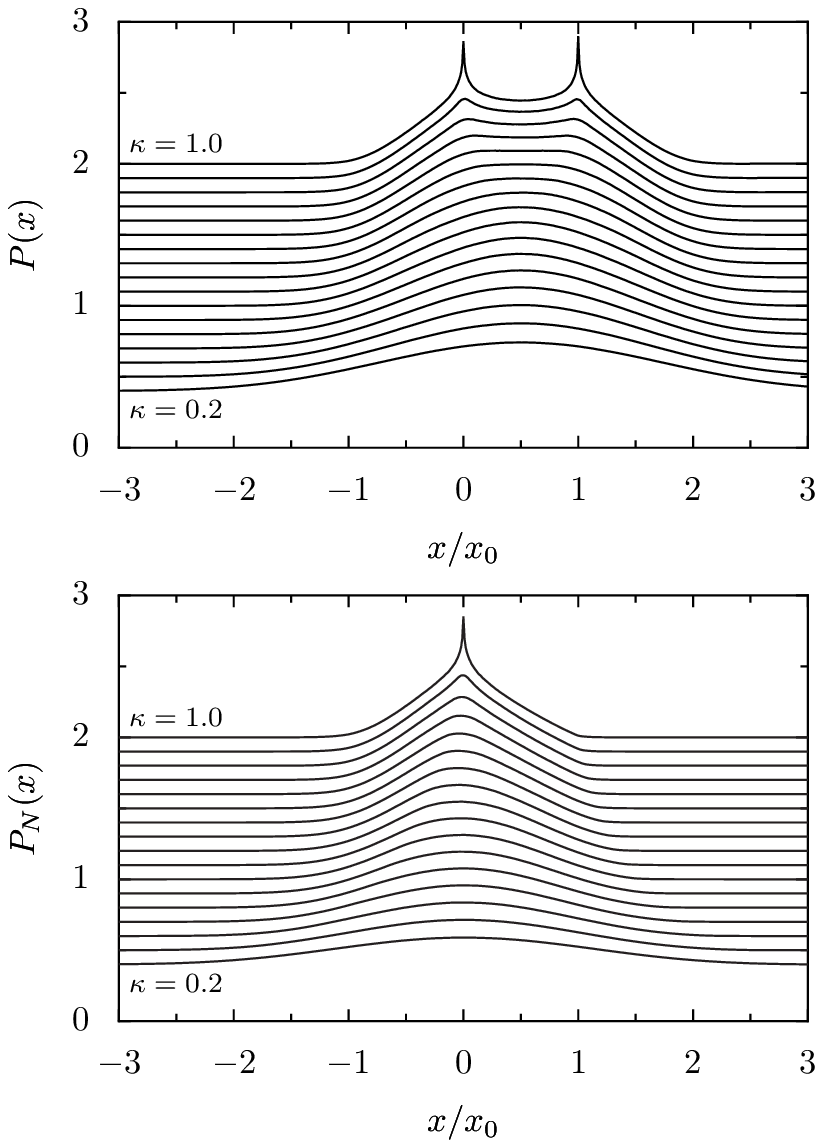}
\caption{Upper panel: Total probability distribution
$P(x)=P_N(x)+P_{N+1}(x)$ of the oscillator's position for different
values of the coupling constant $\kappa$ (defined in
Eq.~(\ref{kappa})). Lower panel: Probability distribution $P_N (x)$ to
find the oscillator at position $x$ if the SET is in charge state
$N$. $P_{N+1}(x)$ can be obtained by the symmetry relation
$P_{N+1}(x)=P_N(1/2-x)$.  In both panels, lines are shifted for
clarity by $2\kappa$, and the difference between neighboring curves is
$\Delta \kappa = 0.05$. All calculations were done at $\epsilon=0.3$
and at the charge-degeneracy point. For the definition of $\epsilon$,
see Eq.~(\ref{epsilon}).}
\label{probx-kappa}
\end{center}
\end{figure}

As can be seen from Fig.~\ref{probx-kappa}, as $\kappa$ is increased,
the stationary position probability distribution evolves continuously
from the weak-coupling Gaussian form to a distribution showing two
sharp peaks at $x=0$ and $x=x_0$ in the limit where $\kappa=1$. This
evolution is the result of a sharpening of each of the two
subdistributions $P_{N(N+1)}(x)$ around their equilibrium position
when $\kappa$ is increased, allowing one to resolve the two
subdistributions individually. This should not only be seen as natural
consequence of the fact that the typical distance $x_0$ scales like
$\sqrt{\kappa}$. In fact, the main cause of the appearance of the two
sharp peaks is that small amplitude oscillations about each of the two
equilibrium points become very stable when $\kappa$ is increased.

We also note that the qualitative shape of each subdistribution evolve
when $\kappa$ is increased. While at low coupling the subdistribution
$P_N(x)$ (resp. $P_{N+1}(x)$) is symmetric about $x=0$
(resp. $x=x_0$), this is not the case for $\kappa \gtrsim 0.4$. This
asymmetry arises only at higher coupling since for low $\kappa$, the
probability to find the oscillator at $x< x^R$ or $x>x^L$ is
negligible. When $\kappa \gtrsim 0.4$, the probability of the
oscillator being in a region transport is forbidden becomes
important. Symmetry breaking arises since these regions are located
only on one side of each equilibrium point.

\begin{figure}
\begin{center}
\includegraphics[scale=0.40]{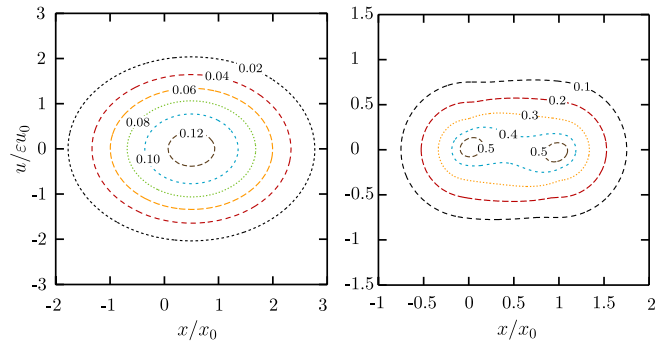}
\caption{Two-dimensional phase-space distributions 
$P(x,u)=P_N(x,u)+P_{N+1}(x,u)$, for
  $\kappa=0.2$ (left panel) and $\kappa=0.9$ (right panel).}
\label{phase_space}
\end{center}
\end{figure}

Finally, we note that the important changes in $P_{N(N+1)} (x)$ that
accompany a variation in $\kappa$ are also seen in the stationary
velocity subdistributions $P_{N(N+1)} (u) = \int du P_{N(N+1)} (x,u)$, that approximatively
follow $P_N ( u/\epsilon u_0) \simeq P_N(x/x_0)$ and $P_{N+1} (
u/\epsilon u_0) \simeq P_{N+1}((x-x_0)/x_0)$, where $u_0 = x_0 /\tau_t$ is the
typical velocity scale in the problem. This can be seen in the
two-dimensional phase-space distributions shown in Fig.~\ref{phase_space}, where both $P(u)$ and $P(x)$ are shown to become more peaked when $\kappa$ is increased. 

For $\kappa > 1$, our numerical investigations show that the current
is strongly suppressed, rendering the intrinsic damping mechanism
discussed at the beginning of this section ineffective. In this case,
the system cannot be characterized by a steady-state probability
distribution, and our model is not appropriate.  Therefore, we only
studied the parameter range $\kappa \le 1$.

A similar reasoning applies to Coulomb-blockade region, where the
damping of the oscillator's motion is severely suppressed.
However, numerically finding a steady-state solution close to the
degeneracy point is possible. 

\section{Average current}
\label{average_current}

The average current flowing through the SET is closely tied to the
oscillator's position distributions via the position-dependent
tunneling rates. Consequently, one expects that the deviations from
the weak-coupling behavior observed in $P(x)=P_N(x)+P_{N+1}(x)$ would
affect the current characteristics when $\kappa$ is increased.

Just like in the previous section, we focus on the degeneracy point
where the average charge state of the island is $N+1/2$. At this
point, the weak-coupling theory predicts\cite{armour04a} that the
current decreases linearly with increasing $\kappa$: $I=
\frac{1-\kappa}{4} (e/\tau_t)$.  This decrease in the current can be
explained in a qualitative way by the reduction of the overlap of the
distributions $P_N(x)$ and $P_{N+1}(x)$ as the coupling is increased,
each distribution becoming more localized around its equilibrium
point, see Fig.~\ref{probx-kappa}.
\begin{figure}
\begin{center}
\includegraphics[scale=1]{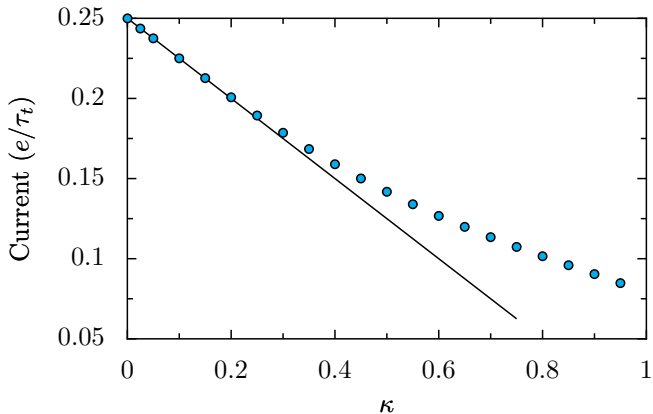}
\caption{Current (in units of $e/\tau_t$) as a function of $\kappa$ at
the degeneracy point $\langle P_N \rangle = \langle P_{N+1} \rangle$,
for $\epsilon=0.3$.  The dots are the results of the Monte Carlo
calculation and the solid line is the analytic form found within the
weak-coupling approximation.}
\label{current vs kappa 1}
\end{center}
\end{figure}

Figure \ref{current vs kappa 1} shows the average current as a
function of $\kappa$. Like in the weak-coupling limit, the
localization of the different probability distributions around its
equilibrium points leads to an overall decrease in the current when
the coupling grows stronger. For $\kappa \gtrsim 0.3$, however, we see
that the numerical results deviate from the weak-coupling behavior:
for stronger coupling the current is higher than the weak-coupling
result.  This can be explained using the rates given by
Eq.~(\ref{approximations}). When either $\int_{x^L} ^{\infty} dx\,
P_{N+1}$ or $\int_{-\infty} ^{x^R} dx \, P_{N}$ is not negligible,
these rates allow unphysical backward currents that are not present in
the full master equation.  For example, a point located at $x > x^L$
in the steady-state probability distribution $P_{N+1} (x,u)$ would
contribute negatively to the average current when using the rates
calculated within the weak-coupling approximation while it would not
contribute to the current when taking into account the full expression
for the rates given in Eq.~(\ref{rates}).

Over the range of frequencies that we studied numerically ($0.1
\leqslant \epsilon \leqslant 0.4$), the current was found to be
practically independent of $\epsilon$ for all but the strongest
couplings ($\kappa \gtrsim 0.8$).  For instance, at $\kappa = 0.9$,
the difference between the calculated currents at $\epsilon=0.1$ and
$\epsilon=0.4$ is on the order of a few percent.

\section{Noise and higher cumulants in the static regime}
\label{noise_higher_moments}
Originally, the interest in SETs was motivated by the suppression of
the current in the Coulomb-blockade regime and the high sensitivity of
the current to small variations of the gate voltage. However, it is
clear that a complete description of the transport processes in these
devices requires not only knowledge of the current, but also of
current-current correlations like e.g., the current
noise.\cite{hershfield93a,korotkov94}  Recently, higher-order
correlations have also been studied both theoretically and
experimentally in nanoscale devices, in the framework of full counting
statistics (FCS) (see Ref.~\onlinecite{nazarov03a} for a collection of
articles on this topic, and Ref.~\onlinecite{bagrets03a,belzig} for a
description of FCS in the context of transport through SETs).  The FCS
approach consists in studying the probability distribution 
$\mathcal{P}_n (t_0)$ that $n$ electrons are transferred through one
lead of the SET within a time period $t_0$, in the limit where
$t_0$ is by far the longest time scale in the problem. 
The full information about the transport properties is 
contained in the cumulants of this distribution function, the first
three of which are given by the average $\mu_1 = \langle n\rangle$,
the variance $\mu_2 = \langle n^2\rangle-\langle n\rangle^2$, and the
skewness $\mu_3=\langle (n-\langle n\rangle)^3\rangle$ that measures
the asymmetry of the distribution.  For example, the current $I=e\langle
n\rangle/t_0$ is proportional to the mean of this distribution, while
the zero-frequency shot noise power $S(0)=2e^2\mu_2/t_0$ is determined
by its second cumulant.

In this section, we study in detail the second and third cumulants of
the probability distribution function $\mathcal{P}_n(t_0)$ in the case of
a coupled SET-nanomechanical system.

\subsection{Weak-coupling case}
It is instructive to start by considering the weak-coupling case
$\kappa \ll 1$, since in this regime we can calculate the noise and
higher cumulants without resorting to Monte Carlo simulations by
solving directly for $\langle n^i (t_0) \rangle$ in the long-time
limit ($t_0 \gamma \gg 1$). In this section, we generalize the work
that was done in Ref.~\onlinecite{armour04b} where a method to
calculate the current-noise using the moments of the steady-state
probability distribution $P_{N(N+1)}(x,u)$ of the oscillator in phase
space was described. In this approach, the current-noise is calculated
from the solution of the equation of motion of $\langle n^2 (t)
\rangle$, the average square of the number of charges that went through a
lead in a time $t$. Here, we extend this method for the calculation of
higher cumulants by deriving the equation of motion for the general
quantity $\langle n^m (t) \rangle$ from which the $m-$th cumulant can
be extracted.

To proceed, we write a master equation for the probability
$P^n_{N(N+1)}(x,u;t)$ to find,
at time $t$, the oscillator at position $x$ with velocity $u$, the
island being in charge state $N(N+1)$, and
that $n$ charges having passed through a lead of the SET in the
interval $[0;t]$. We will again make the assumptions leading to
Eq.~(\ref{approximations}). Considering for definitiveness the left
lead, at zero temperature and neglecting any extrinsic damping, one
finds \cite{armour04b}

\begin{subequations}\label{noise_master_equation}
\begin{align}
\frac{\partial }{\partial t} P^n_N (x,u;t)&= \omega_0^2 x
\frac{\partial} 
{\partial u}P^n_N(x,u;t)- u \frac{\partial }{\partial x} P^n_N(x,u;t) 
\nonumber\\
&+\Gamma _L^+ (x) P^{n-1}_{N+1}(x,u;t)- \Gamma _R^+ (x)P^n_{N}(x,u;t)\; ,
\nonumber\\
\label{noise_master_equation_npn}
\end{align}
\begin{align}
\frac{\partial }{\partial t} P^n_{N+1} (x,u;t)&= \omega_0^2 (x-x_0)  
\frac{\partial} {\partial u}P^n_{N+1}(x,u;t)\nonumber\\
&- u \frac{\partial }{\partial x} P^n_{N+1}(x,u;t) \nonumber\\
&-\Gamma _L^+ (x)  P^n_{N+1}(x,u;t)+ \Gamma _R^+ (x)
P^n_{N}(x,u;t)\; .\nonumber\\
\label{noise_master_equation_npnp1}
\end{align}
\end{subequations}
where the rates are taken from Eq.~(\ref{approximations}). Defining the coupled
moments $\langle x^j u^k n^m \rangle$ and $\langle x^j u^k n^m
\rangle_{N+1}$ as
\begin{subequations}
\begin{align}
\langle n^m x^j u^k \rangle &= \nonumber \\ \sum_n n^m &
\int \, du \int \, dx \, x^j u^k
\left [ P^n_N (x,u;\tau)+ P^n_{N+1} (x,u;\tau) \right] \;,
\end{align}
\begin{align}
 \langle n^m x^j u^k \rangle_{N+1} &= \sum_n n^m 
\int \, du \int \, dx \, x^j u^k  P^n_{N+1} (x,u;\tau)\;
,\nonumber\\
\end{align}
\end{subequations}
one can calculate the equation of motion for these quantities using Eq. (\ref{noise_master_equation}). With $x$ in units of $x_0$ and $u$ in units of $u_0$, one finds
\begin{subequations}\label{label FCS}
\begin{multline}
\frac{d}{d\tau} \langle x^j 
u^k n^m \rangle _{N+1} = \\
-k \epsilon^2 \left[ \langle x^{j+1} u^{k-1}
  n^m\rangle_{N+1} -  \langle x^j u^{k-1} n^m
  \rangle_{N+1} \right]\\
 + j \langle x^{j-1}u^{k+1} n^m\rangle_{N+1} 
- \langle x^j u^k n^m \rangle_{N+1} \\
+ \Delta_R \langle x^j u^kn^m \rangle + 
\kappa \langle x^{j+1}u^kn^m\rangle\; ,
\label{label FCS np1}
\end{multline}
\begin{multline}
\frac{d}{d\tau}\langle x^j u^k n^m
\rangle = \\
-k \epsilon^2 \left[ \langle x^{j+1} u^{k-1 } n^m
  \rangle -  \langle x^j u^{k-1} n^m \rangle_{N+1}
  \right]\\
 + j \langle x^{j-1}u ^{k+1}n^m
\rangle\\
 + \sum_{i=0} ^{m-1} \binom {m} {i} \left[ \Delta_L \langle
  x^j u^k n^i \rangle_{N+1} -\kappa \langle 
x^{j+1}u^k n^i\rangle_{N+1} \right]\; .
\label{label FCS ntot}
\end{multline}
\end{subequations}
Here, averages that are independent of $n$ (averages of the form
$\langle x^j u^k n^0 \rangle$) are time-independent and can be
evaluated in the stationary limit, i.e., 
Eqs.~(\ref{label FCS np1}-\ref{label FCS ntot}) can be used to 
generate a closed linear
system of equations.\cite{footnote2} The terms $\langle x^j u^k
\rangle$ of order $j+k=c$ are connected to the terms $\langle x^j u^k
\rangle _{N+1}$ of order $j+k=c-1$. This means that to solve for a
moment $\langle x^j u^k \rangle$, we must use the $c+1$ equations of
the type of Eq.~(\ref{label FCS ntot}) where $j+k=c$ and the $c$
equations of the type of Eq.~(\ref{label FCS np1}) where $j+k=c-1$. 
This method can be used to calculate any moment of the form 
$\langle x^j u^k\rangle$ and $\langle x^j u^k\rangle_{N+1}$.
Knowledge of $\langle x^j u^k \rangle$ enables one to calculate
the long-time behavior of the coupled moments of the charge and
oscillator's position in phase space $\langle x^{j'} u^{k'} n^i \rangle$,
and thus the $i$-th moment $\langle n^i \rangle$ of $\mathcal{P}_n$.

The ratio of the zero-frequency shot noise power and the average
current (times $2e$), or equivalently, the ratio of the second
and first cumulants of $\mathcal{P}_n$ is called Fano factor and is
readily calculated using this approach. Since it shows a complex
dependence on the coefficients $\Delta_L$, $\Delta_R$, and on the parameters
$\kappa$ and $\epsilon$, it is convenient to expand the result in
powers of $\kappa$. Introducing a parameter $\alpha$ defined via $
\alpha= \Delta_L- (1+\kappa)/2$ (or, equivalently, $\alpha =-\Delta_R
+ (1-\kappa)/2$) that measures the difference between $\Delta_L$ and
its value at the degeneracy point, one can write down the Fano factor
in a way that underlines its symmetry with respect to this point:
\begin{align}
\begin{split}
\frac{S(0)}{2 e I}& = \frac{1}{2} + 2 \alpha^2 + 4 \alpha^2 \kappa +
6 \alpha^2 \kappa^2
\\&+\left( \frac{1}{2} - 2 \alpha^2 \right) \frac{\kappa}{\epsilon^2}
-  \left( \frac{1}{2} + 2 \alpha^2 \right) \frac{\kappa^2}{\epsilon^2}
+\mathcal{O}(\kappa ^3)\; .
\end{split}
\end{align}
For $\epsilon \ll 1$,
the Fano factor is dominated by the term $\sim\kappa/\epsilon^2$, like
in the case where one considers a system of two SETs coupled by an
oscillator.\cite{rodrigues05a} Finally, we note that current
conservation implies that the Fano factor is identical in both
leads.\cite{blanter00a}

\begin{figure}
\begin{center}
\includegraphics[scale=1]{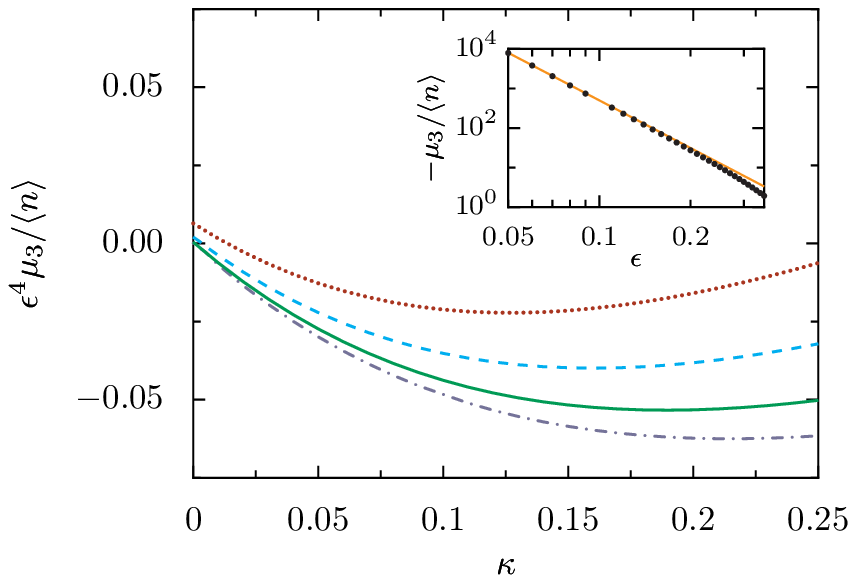}
\caption{Normalized third cumulant as a function of $\kappa$ for
different values of $\epsilon$, as calculated within the weak-coupling
approximation and scaled by $\epsilon^4$. Dotted line: $\epsilon=0.4$,
dashed line: $\epsilon=0.3$, solid line: $\epsilon=0.2$, dash-dotted
line: $\epsilon=0.1$. The inset shows the $\epsilon$-dependence of the
normalized third cumulant at $\kappa=0.1$ (symbols), as calculated
within the weak-coupling approximation. The solid line is a fit to a
power law $\sim \epsilon^{-4}$. These results were obtained by
integrating the equation of motion for $\langle n^3(t)\rangle$
following from Eq.~(\ref{label FCS}).}
\label{wcthirdmoment}
\end{center}
\end{figure}

Equation \ref{label FCS} is one of the main result of our article, as it 
allows for the calculation of higher cumulants of the
current by integrating the equation of motion for the moments of the
form $\langle x^j u^k n^m (t) \rangle$ with $m> 0$. For example, we
calculated the normalized third cumulant $\mu_3 / \langle n \rangle$
of $\mathcal{P}_n (t_0)$. The results are presented in
Fig.~\ref{wcthirdmoment}. We stress that these results have been
obtained by integrating the equation motion for $\langle n^3 (t)
\rangle$ valid in the weak-coupling regime and not via a Monte Carlo
simulation.  Starting from the value $1/4$ at
$\kappa=0$,\cite{bagrets03a} the normalized third cumulant decreases
rapidly when $\kappa$ is increased. On further increase of $\kappa$,
it reaches a minimum at an $\epsilon$-dependent value of $\kappa$. The
inset of Fig.~\ref{wcthirdmoment} shows that the leading contribution
to the normalized third cumulant in the weak-coupling regime is of the
form $\epsilon^{-4}$. As a consequence, we note that the asymmetry of
the probability distribution $\mathcal{P}_n$ that is determined by
$\mu_3$, can effectively be tuned by changing the frequency of the
oscillator or $\tau_t$. The scaled quantity $\epsilon^4 \mu_3 /
\langle n \rangle$ shows contributions of higher-order corrections in
$\epsilon$ to the normalized third cumulant.

\subsection{Higher coupling}
It is unfortunately not straightforward to generalize the previously
described method for calculating the cumulants of
$\mathcal{P}_n$ outside the weak-coupling regime. The presence of
$x$-dependent Fermi functions in the tunneling rates as well as the
possibility of charge flow against the direction set by the bias
voltage due to the position of the oscillator gives rise to a system of
equations that is not closed and cannot be solved analytically. Even
if we neglect transport against the dominant direction of the current
$\Gamma_j^- (x) \simeq 0$, but keep the position
dependence of the Fermi distributions in $\Gamma_j ^+ (x)$, it is
still not possible to derive a system of equations coupling only
objects of the form $\langle x^j u^k n^m \rangle$. Therefore, we will
use a numerical approach to evaluate the cumulants of
$\mathcal{P}_n$.

Indeed, the Monte Carlo method described in Appendix \ref{MCmethod}
can be used to measure the FCS of electron transport in the same way
as it can be done in experiments.\cite{gustavsson06a} A very long
Monte Carlo run is divided into intervals of duration $t_0 \gg
\gamma^{-1}$, here, $\gamma=\kappa\epsilon^2$ is the damping constant
defined at the beginning of Sec. \ref{dynamics_oscillator};
$\gamma^{-1}$ is the longest intrinsic time scale of the problem. By
counting the number of charges going through one lead during each
interval, one can reconstruct the probability distribution
$\mathcal{P}_n (t_0)$, and from it calculate the cumulants.

\begin{figure}
\begin{center}
\includegraphics[scale=1]{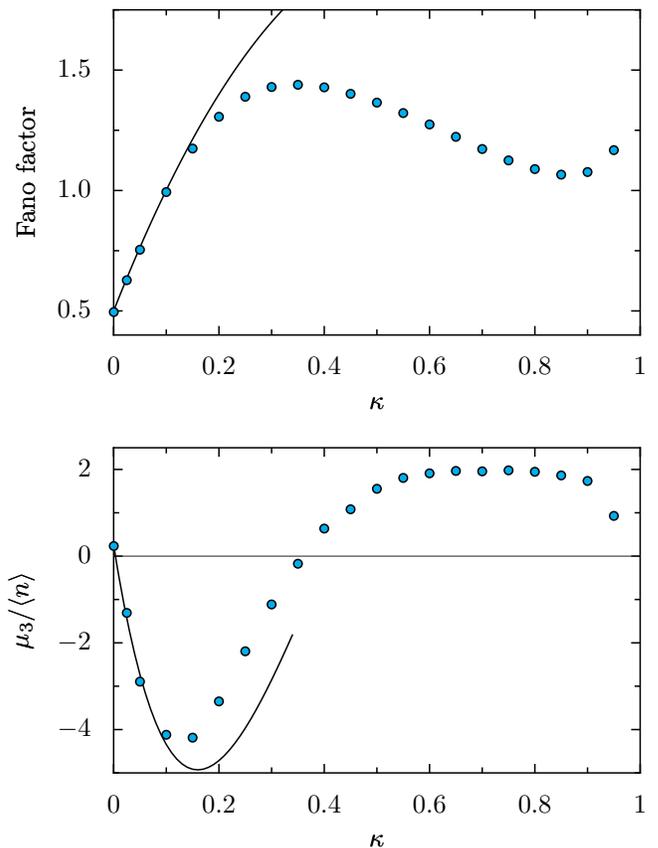}
\caption{Upper panel: Fano factor as a function of $\kappa$ at the
degeneracy point $\langle P_N \rangle = \langle P_{N+1} \rangle$. The
dots are the results of the numerical calculation and the solid line
is the analytic form found within the weak-coupling
approximation. Lower panel: Normalized third cumulant $\mu_3$ of the
probability distribution $\mathcal{P}_ n$. For both panels,
$\epsilon=0.3$.}
\label{figurefanothirdm}
\end{center}
\end{figure}

We study current correlations at the charge degeneracy point, where
the average charge state of the island is $N+1/2$. The top panel of
Fig.~\ref{figurefanothirdm} compares the weak-coupling Fano factor to
the numerical Monte Carlo results for different values of the
coupling parameter $\kappa$. Naturally, for $\kappa \lesssim 0.2$, the
agreement between the numerical results and those obtained
analytically is very good. Beyond this point, the numerically
calculated Fano factor shows an interesting non-monotonic behavior,
with a maximum at $\kappa \sim 0.35$ and a minimum at $\kappa \sim
0.85$.  The lower panel of Fig.~\ref{figurefanothirdm} also shows the
evolution of the normalized third cumulant of $\mathcal{P}_n$, giving
the asymmetry of this probability distribution about its mean $\langle
n \rangle$. Starting from the $\kappa=0$ value of $1/4$ derived for
a simple SET device, our results show that this quantity is, in the
weak-coupling limit, very sensitive to variations of $\kappa$. Indeed,
the normalized third cumulant changes sign twice in
the region $\kappa \lesssim 0.35$, reaching a maximum value
approximatively in the middle of this region. This contrasts with the
strong-coupling behavior: $\mu_3 / \langle n \rangle$ stays
practically constant for $0.5 \lesssim \kappa \lesssim 0.9$.
 
\begin{figure}
\begin{center}
\includegraphics[scale=1]{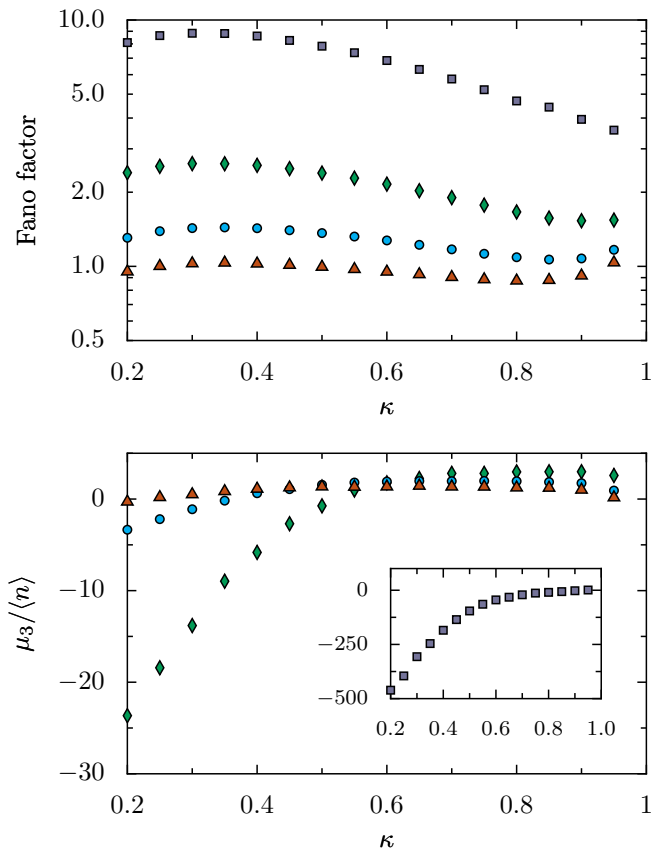}
\caption{Fano factor (upper panel) and normalized third cumulant
(lower panel and inset) as a function of $\kappa$ for different values
of $\epsilon$: $\epsilon=0.1$ (squares in the upper panel and in the
inset), $\epsilon=0.2$ (diamonds), $\epsilon=0.3$ (circles), and
$\epsilon=0.4$ (triangles). Note the logarithmic $y$-axis in the upper
panel.}
\label{fano_eps}
\end{center}
\end{figure}

We will now address the question how the previous results are modified
when changing the frequency of the oscillator. Figure~\ref{fano_eps}
shows the dependence of the Fano factor and the normalized third
cumulant as a function of $\kappa$ for different values of
$\epsilon$. First, we note that the actual value of the Fano factor is
increased dramatically for lower oscillator frequencies, as expected
from the term $\sim \kappa / \epsilon^2$ that dominates in the
low-frequency regime. In the weak-coupling region ($\kappa \lesssim
0.3$), the magnitude of the normalized third cumulant is also heavily
affected by a change in frequency, in agreement with the weak-coupling
leading order dependence $\sim \epsilon^{-4}$. Despite these major
changes in magnitude of both the Fano factor and the normalized third
cumulant, the overall qualitative effect of an increase in coupling
does not seem to depend heavily on $\epsilon$, in the frequency range
we investigated. In particular, the position of the maximum in the
Fano factor remains constant. Also, the normalized third cumulant
always shows a change of sign, albeit at an $\epsilon$-dependent value
of $\kappa$, and goes to a positive for $\kappa \to 1$. 
Remarkably, the value of the normalized third cumulant is much
less sensitive to $\epsilon$ in the strong-coupling regime.
This might be the signature of a transport regime in the $\kappa \to 1 
$ region that is radically different of the one found for $\kappa  
\simeq 0.2$.

\section{Frequency-dependent noise}
\label{frequency-dependent_noise}
In systems that exhibit internal dynamics like the one we study, it is
especially interesting to look at the frequency-dependence of the
current-current correlations. In Ref.~\onlinecite{armour04b}, the
frequency-dependent noise $S(\omega)$ of a SET weakly coupled to a
nanomechanical oscillator was thoroughly studied. It was found that,
the noise spectrum shows only two peaks at finite frequency at $\omega_0'$
and $2\omega_0'$, where $\omega_0^\prime = \omega_0\sqrt{1-\kappa} $
is the effective frequency of the damped harmonic oscillator. In this
section, we extend the work of Ref.~\onlinecite{armour04b} by studying
the frequency-dependent noise power $S(\omega)$ in the strong-coupling
regime ($0.2 \lesssim \kappa< 1$).

In order to calculate the frequency-dependent noise using our
Monte Carlo method, we follow the approach developed by
MacDonald\cite{macdonald62a,korotkov94} that was used recently to
study the noise properties of mesoscopic systems, including coupled
SET-nanomechanical systems in the weak-coupling regime. In general,
the current-noise power at junction $a$ is defined as the Fourier
transform of the current autocorrelation function $K_{i,i}$ at
junction $i$,
\begin{equation}
S_{i,i} (\omega) = 2 \int_{-\infty} ^\infty d\tau \, \cos(\omega \tau) 
K_{i,i} (\tau)\; ,
\end{equation}
where
\begin{equation}
K_{i,i} (\tau) = \langle I_i (t+\tau) I_i (t) \rangle- \langle I_i
\rangle ^2\; .
\end{equation}
To proceed with the MacDonald approach, $I_i$, and therefore $I_i -
\langle I_i \rangle$, must be assumed to be statistically fluctuating
variables, such that the autocorrelation function $K_{i,i}$ is
independent of $t$. In this case, the MacDonald formula 
relates the fluctuation $\delta n$ about the average of the number of
charges $n$ that went through a junction in time $\tau$,
\begin{equation}
\delta n_i(\tau) = n_i(\tau) - \langle I_i
\rangle \tau = \int _t ^{t+\tau} d t' \, \big( I_i
(t') - \langle I_i \rangle \big) \; ,
\end{equation}
to the current-noise power via
\begin{equation}
S_{i,i}(\omega) = 2 \omega \int _0 ^\infty \, d \tau \, \sin(\omega
\tau) \, \frac{\partial}{\partial \tau} \langle   ( \delta n_i(\tau))
^2 \rangle\; ,
\end{equation}
where $\langle ( \delta n_i(\tau)) ^2 \rangle = \langle n_i^2 (\tau)
\rangle - \langle I_i \rangle^2 \tau^2$. Since $\langle n^2 (\tau)
\rangle$ and $\langle I_i \rangle$ are easily accessible through the
Monte Carlo simulation, $S(\omega)$ can be calculated by taking a
numerical time-derivative of $\langle ( \delta n_i(\tau)) ^2 \rangle$
and then evaluating the Fourier sine-transform. Note that we consider
only the particle current fluctuations here. The electrical current
noise at finite frequencies includes a contribution from displacement
currents, which depend on the capacitive couplings between the island
and the leads.\cite{blanter00a} Since we assume that our frequencies
of interest are much smaller than the relevant RC-frequencies, we can
neglect the displacement currents here, see e.g. the discussion in
Ref.~\onlinecite{cottet04:prl,cottet04:prb}.

\begin{figure}
\begin{center}
\includegraphics[scale=0.98]{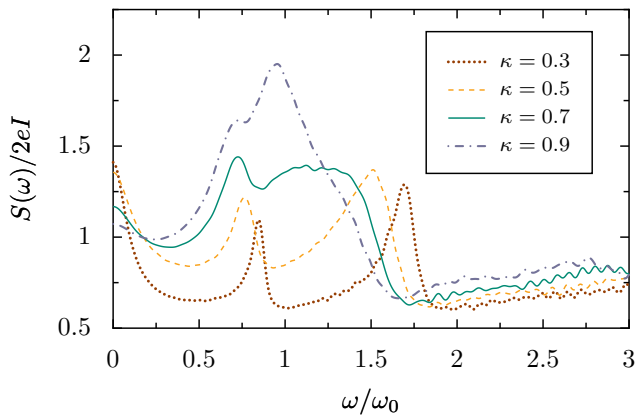} 
\caption{Frequency-dependent noise power beyond the weak-coupling
approximation. For each curve, the SET is tuned to the
charge-degeneracy point and $\epsilon=0.3$.}
\label{freqdephc}
\end{center}
\end{figure}

The results of the Monte Carlo simulation are shown in
Fig.~\ref{freqdephc}. Like in the weak-coupling case, $S(\omega)$
shows two main finite-frequency features.  Surprisingly, even for
strong coupling, we do not find any features for frequencies higher
than $2\omega_0$.  We find a low-frequency peak that defines the
frequency $\omega_0^\prime$ (which will be different from the
weak-coupling prediction $\omega_0\sqrt{1-\kappa}$ in the general
case). The second feature evolves from the peak located at $2 \omega_0
^\prime$ predicted by the weak-coupling theory. While both peaks are
considerably broadened by an increase in the coupling strength, their
respective shapes evolve in a qualitatively different way. Whereas the
first peak is shifted in absolute frequency, the second peak broadens
in a very asymmetric way, with much of the weight shifting to lower
frequencies. The slope of its left shoulder decreases with increasing
$\kappa$ until it forms a plateau at around $\kappa \sim 0.7$. On
increasing $\kappa$ even further, the two peaks merge, leading to
super-Poissonian frequency-dependent noise throughout the frequency
range $\omega < 1.5 \omega_0$.

\begin{figure}
\begin{center}
\includegraphics[scale=1]{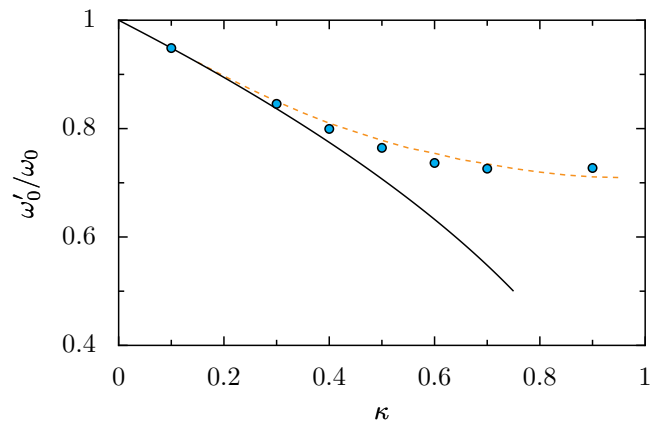}
\caption{Position $\omega_0^\prime$ of the first peak in the
frequency-dependent noise power as a function of $\kappa$. The solid
line gives the weak-coupling prediction $\omega_0\sqrt{1-\kappa}$, the
data points are the numerical Monte Carlo results, and the dashed line
was obtained using an effective damping constant, see text.}
\label{first_peak}
\end{center}
\end{figure}

Figure~\ref{first_peak} shows the position of the maxima of the first
peak in the frequency-dependent noise power as a function of $\kappa$.
By comparing the position of the first peak extracted from the curves
shown in Fig.~\ref{freqdephc} (data points in Fig.~\ref{first_peak}) to
the weak-coupling prediction $\omega_0^\prime =
\omega_0\sqrt{1-\kappa}$ (solid line in Fig.~\ref{first_peak}), we
find quantitative agreement only for $\kappa \lesssim 0.2$.
Beyond this point, the ratio $\omega_0^\prime / \omega_0$ still
decreases, albeit slowly, when $\kappa$ is increased. It reaches a
saturation value $\omega_0^\prime \sim 0.7 \omega_0 $ for $\kappa
\gtrsim 0.7$.

This behavior can be understood by interpreting the frequency shift in
terms of an effective damping mechanism caused by electron
tunneling. Since there is no damping without current, the natural
modification of the weak-coupling damping constant $\gamma = \kappa
\epsilon^2 $ in the strong-coupling regime is to renormalize the
weak-coupling result by the probability $P^*$ to find the oscillator
in a position where in principle current is allowed, i.e., for $x^R
<x$ and charge state $N$, or $x < x^L$ and charge state $N+1$.
Defining a renormalized damping constant $\gamma ^{sc} = P^* \kappa
\epsilon^2$, it is possible to estimate the position of the first peak
as a function of $\kappa$ using values of $P^*$ extracted from curves
presented in Fig.~\ref{probx-kappa}. The result is shown as the dashed
curve in Fig.~\ref{first_peak} and agrees with the data points
obtained by the Monte Carlo method in a quantitative way.

\section{Conclusion}
In this paper, we have studied the strong-coupling limit of a SET
transistor coupled to a classical harmonic oscillator. We have used a
combination of a master-equation approach and a numerical Monte Carlo
procedure to calculate the position distribution of the oscillator,
the electrical current, and the zero-frequency noise in both the
weak-coupling and strong-coupling regime. With increasing coupling, we
found that the position distribution of the oscillator evolves from a
broad Gaussian to a a function sharply peaked around each of the
charge-state dependent equilibrium positions of the oscillator. We
found that the average current in the strong-coupling regime is higher
than the value predicted by the weak- coupling theory and that the
Fano factor varies in a non-monotonous fashion when coupling is
increased.  We have generalized the weak-coupling theory to allow
the calculation of higher cumulants of the current, and have presented
results for the third cumulant. In the weak-coupling regime, the third
cumulant was found to depend strongly on the frequency of the
oscillator, whereas in the strong-coupling regime it becomes
practically independent of this parameter.  We have also studied the
frequency-dependent transport noise. Even in the strong-coupling
regime, there are no peaks for frequencies higher than $2\omega_0$,
and the two peaks found in the weak-coupling limit merge on increasing
the coupling strength.  Finally, we introduced a generalized
expression connecting the damping rate in the strong-coupling regime
with the other parameters of our model and used it to understand the
evolution of the oscillator's damping-renormalized frequency as a
function of coupling.

\section{Acknowledgments}
We would like to thank J. Lehmann and M. Vanevic, and particularly
A.~D. Armour for interesting discussions and correspondence.  This
work was financially supported by the Natural Sciences and Engineering
Research Council of Canada, the Fonds Qu\'eb\'ecois de la Recherche
sur la Nature et les Technologies, the SFB 513 Nanostructures at
Surfaces and Interfaces of the DFG, the Swiss NSF, and the NCCR
Nanoscience.

\appendix
\section{Details of the Monte Carlo approach used}
\label{MCmethod}
Monte Carlo methods have been used for a long time to calculate
numerically the transport properties of mesoscopic systems like
SETs.\cite{amman89a} When dealing with a simple SET system, the idea
of the Monte Carlo approach is to solve the master equation for the
charge states of the SET by discretizing time into small intervals and
allowing charge transfer to and from the dot with a probability that
is proportional to the tunneling rates and the time interval between
two attempts.

If the SET is coupled to a harmonic oscillator, we can proceed in a
similar way, by considering charge transfer attempts at a finite
frequency $(\Delta \tau_t)^{-1}$, where $\Delta \ll 1$ is a
dimensionless step size. The success probability for a charge transfer
is calculated from the oscillator's position-dependent instantaneous
rates $\Gamma_i^\sigma(x)$ calculated at the time of the attempt. Between
different attempts, the oscillator's state evolves according to the
classical equation of motion, whose solution depends on the charge
state of the SET, the equilibrium position of the oscillator being
shifted by $x_0$ when the charge state is changed from $N \to N+1$ or
by $-x_0$ in the opposite case. At the beginning of each run, the
state of the system is determined randomly from the stationary
probability distributions $P_N (x,u)$ and $P_{N+1} (x,u)$. In
practice, this can be implemented by using the final state of the
$(n-1)$-th Monte Carlo run as the initial state of the $n$-th run.

We consider runs of total time $t_0 \tau_t$, such that each run
consists of $M = t_0 / \Delta $ steps. Both the time scales $t_0
\tau_t$ and $\Delta \tau_t $ are chosen in a way such that increasing
$t_0$ or decreasing $\Delta$ does not affect the value of the
different physical quantities we extract from our calculation. In
practice, this corresponds to choosing $\Delta < 0.02$ and $t_0
\tau_t$ an order of magnitude greater than the typical damping time
$1/\gamma$. A consequence of this last constraint is that the
Monte Carlo approach is particularly useful in the strong-coupling
regime, where the number of steps $M$ per run can be kept relatively
small, allowing for more runs to be made in the same amount of
computer time.

We checked that our code reproduces the analytical results of
Ref.~\onlinecite{bagrets03a} for the dependence of the noise and the
third cumulant on the asymmetry coefficients $\Delta_L - \Delta_R$ in
the $\kappa=0$ limit. Also, the probability distributions $P_N (x,u)$
and $P_{N+1} (x,u)$ that we calculate using the Monte Carlo approach
coincide with the results one finds when solving Eq.~(\ref{master
equation}) on a grid.\cite{armour04a}


\begin{thebibliography}{42}
\expandafter\ifx\csname natexlab\endcsname\relax\def\natexlab#1{#1}\fi
\expandafter\ifx\csname bibnamefont\endcsname\relax
  \def\bibnamefont#1{#1}\fi
\expandafter\ifx\csname bibfnamefont\endcsname\relax
  \def\bibfnamefont#1{#1}\fi
\expandafter\ifx\csname citenamefont\endcsname\relax
  \def\citenamefont#1{#1}\fi
\expandafter\ifx\csname url\endcsname\relax
  \def\url#1{\texttt{#1}}\fi
\expandafter\ifx\csname urlprefix\endcsname\relax\def\urlprefix{URL }\fi
\providecommand{\bibinfo}[2]{#2}
\providecommand{\eprint}[2][]{\url{#2}}

\bibitem[{\citenamefont{Cleland and Roukes}(1998)}]{roukes98}
\bibinfo{author}{\bibfnamefont{A.}~\bibnamefont{Cleland}} \bibnamefont{and}
  \bibinfo{author}{\bibfnamefont{M.}~\bibnamefont{Roukes}},
  \bibinfo{journal}{Nature} \textbf{\bibinfo{volume}{392}},
  \bibinfo{pages}{160} (\bibinfo{year}{1998}).

\bibitem[{\citenamefont{Blencowe}(2004)}]{blencowe04}
\bibinfo{author}{\bibfnamefont{M.~P.} \bibnamefont{Blencowe}},
  \bibinfo{journal}{Phys. Rep.} \textbf{\bibinfo{volume}{395}},
  \bibinfo{pages}{159} (\bibinfo{year}{2004}).

\bibitem[{\citenamefont{Roukes}(2001)}]{roukes01}
\bibinfo{author}{\bibfnamefont{M.}~\bibnamefont{Roukes}},
  \bibinfo{journal}{Phys. World} \textbf{\bibinfo{volume}{14}},
  \bibinfo{pages}{25} (\bibinfo{year}{2001}).

\bibitem[{\citenamefont{LaHaye et~al.}(2004)\citenamefont{LaHaye, Buu,
  Camarota, and Schwab}}]{schwab04}
\bibinfo{author}{\bibfnamefont{M.}~\bibnamefont{LaHaye}},
  \bibinfo{author}{\bibfnamefont{O.}~\bibnamefont{Buu}},
  \bibinfo{author}{\bibfnamefont{B.}~\bibnamefont{Camarota}}, \bibnamefont{and}
  \bibinfo{author}{\bibfnamefont{K.}~\bibnamefont{Schwab}},
  \bibinfo{journal}{Science} \textbf{\bibinfo{volume}{304}},
  \bibinfo{pages}{74} (\bibinfo{year}{2004}).

\bibitem[{\citenamefont{Gorelik et~al.}(1998)\citenamefont{Gorelik, Isacsson,
  Voinova, Kasemo, Shekhter, and Jonson}}]{gorelik98}
\bibinfo{author}{\bibfnamefont{L.}~\bibnamefont{Gorelik}},
  \bibinfo{author}{\bibfnamefont{A.}~\bibnamefont{Isacsson}},
  \bibinfo{author}{\bibfnamefont{M.}~\bibnamefont{Voinova}},
  \bibinfo{author}{\bibfnamefont{B.}~\bibnamefont{Kasemo}},
  \bibinfo{author}{\bibfnamefont{R.}~\bibnamefont{Shekhter}}, \bibnamefont{and}
  \bibinfo{author}{\bibfnamefont{M.}~\bibnamefont{Jonson}},
  \bibinfo{journal}{\prl} \textbf{\bibinfo{volume}{80}}, \bibinfo{pages}{4526}
  (\bibinfo{year}{1998}).

\bibitem[{\citenamefont{Weiss and Zwerger}(1999)}]{zwerger99}
\bibinfo{author}{\bibfnamefont{C.}~\bibnamefont{Weiss}} \bibnamefont{and}
  \bibinfo{author}{\bibfnamefont{W.}~\bibnamefont{Zwerger}},
  \bibinfo{journal}{Europhys. Lett.} \textbf{\bibinfo{volume}{47}},
  \bibinfo{pages}{97} (\bibinfo{year}{1999}).

\bibitem[{\citenamefont{Erbe et~al.}(2001)\citenamefont{Erbe, Weiss, Zwerger,
  and Blick}}]{blick01}
\bibinfo{author}{\bibfnamefont{A.}~\bibnamefont{Erbe}},
  \bibinfo{author}{\bibfnamefont{C.}~\bibnamefont{Weiss}},
  \bibinfo{author}{\bibfnamefont{W.}~\bibnamefont{Zwerger}}, \bibnamefont{and}
  \bibinfo{author}{\bibfnamefont{R.}~\bibnamefont{Blick}},
  \bibinfo{journal}{\prl} \textbf{\bibinfo{volume}{87}},
  \bibinfo{pages}{096106} (\bibinfo{year}{2001}).

\bibitem[{\citenamefont{Pistolesi}(2004)}]{pistolesi04}
\bibinfo{author}{\bibfnamefont{F.}~\bibnamefont{Pistolesi}},
  \bibinfo{journal}{\prb} \textbf{\bibinfo{volume}{69}},
  \bibinfo{pages}{245409} (\bibinfo{year}{2004}).

\bibitem[{\citenamefont{Isacsson and Nord}(2004)}]{isacsson04}
\bibinfo{author}{\bibfnamefont{A.}~\bibnamefont{Isacsson}} \bibnamefont{and}
  \bibinfo{author}{\bibfnamefont{T.}~\bibnamefont{Nord}},
  \bibinfo{journal}{Europhys. Lett.} \textbf{\bibinfo{volume}{66}},
  \bibinfo{pages}{708} (\bibinfo{year}{2004}).

\bibitem[{\citenamefont{Novotny et~al.}(2004)\citenamefont{Novotny, Donarini,
  Flindt, and Jauho}}]{jauho04}
\bibinfo{author}{\bibfnamefont{T.}~\bibnamefont{Novotny}},
  \bibinfo{author}{\bibfnamefont{A.}~\bibnamefont{Donarini}},
  \bibinfo{author}{\bibfnamefont{C.}~\bibnamefont{Flindt}}, \bibnamefont{and}
  \bibinfo{author}{\bibfnamefont{A.-P.} \bibnamefont{Jauho}},
  \bibinfo{journal}{\prl} \textbf{\bibinfo{volume}{92}},
  \bibinfo{pages}{248302} (\bibinfo{year}{2004}).

\bibitem[{\citenamefont{Pistolesi and Fazio}(2006)}]{pistolesi06}
\bibinfo{author}{\bibfnamefont{F.}~\bibnamefont{Pistolesi}} \bibnamefont{and}
  \bibinfo{author}{\bibfnamefont{R.}~\bibnamefont{Fazio}},
  \bibinfo{journal}{New J. of Phys.} \textbf{\bibinfo{volume}{8}},
  \bibinfo{pages}{113} (\bibinfo{year}{2006}).

\bibitem[{\citenamefont{Fedorets et~al.}(2004)\citenamefont{Fedorets, Gorelik,
  Shekhter, and Jonson}}]{fedorets04}
\bibinfo{author}{\bibfnamefont{D.}~\bibnamefont{Fedorets}},
  \bibinfo{author}{\bibfnamefont{L.~Y.} \bibnamefont{Gorelik}},
  \bibinfo{author}{\bibfnamefont{R.~I.} \bibnamefont{Shekhter}},
  \bibnamefont{and} \bibinfo{author}{\bibfnamefont{M.}~\bibnamefont{Jonson}},
  \bibinfo{journal}{\prl} \textbf{\bibinfo{volume}{92}}, \bibinfo{eid}{166801}
  (\bibinfo{year}{2004}).

\bibitem[{\citenamefont{Fedorets et~al.}(2005)\citenamefont{Fedorets, Gorelik,
  Shekhter, and Jonson}}]{fedorets05}
\bibinfo{author}{\bibfnamefont{D.}~\bibnamefont{Fedorets}},
  \bibinfo{author}{\bibfnamefont{L.~Y.} \bibnamefont{Gorelik}},
  \bibinfo{author}{\bibfnamefont{R.~I.} \bibnamefont{Shekhter}},
  \bibnamefont{and} \bibinfo{author}{\bibfnamefont{M.}~\bibnamefont{Jonson}},
  \bibinfo{journal}{\prl} \textbf{\bibinfo{volume}{95}}, \bibinfo{eid}{057203}
  (\bibinfo{year}{2005}).

\bibitem[{\citenamefont{Armour et~al.}(2004)\citenamefont{Armour, Blencowe, and
  Zhang}}]{armour04a}
\bibinfo{author}{\bibfnamefont{A.~D.} \bibnamefont{Armour}},
  \bibinfo{author}{\bibfnamefont{M.~P.} \bibnamefont{Blencowe}},
  \bibnamefont{and} \bibinfo{author}{\bibfnamefont{Y.}~\bibnamefont{Zhang}},
  \bibinfo{journal}{\prb} \textbf{\bibinfo{volume}{69}},
  \bibinfo{pages}{125313} (\bibinfo{year}{2004}).

\bibitem[{\citenamefont{Armour}(2004)}]{armour04b}
\bibinfo{author}{\bibfnamefont{A.~D.} \bibnamefont{Armour}},
  \bibinfo{journal}{\prb} \textbf{\bibinfo{volume}{70}},
  \bibinfo{pages}{165315} (\bibinfo{year}{2004}).

\bibitem[{\citenamefont{Mozyrsky et~al.}(2004)\citenamefont{Mozyrsky, Martin,
  and Hastings}}]{imartin04}
\bibinfo{author}{\bibfnamefont{D.}~\bibnamefont{Mozyrsky}},
  \bibinfo{author}{\bibfnamefont{I.}~\bibnamefont{Martin}}, \bibnamefont{and}
  \bibinfo{author}{\bibfnamefont{M.~B.} \bibnamefont{Hastings}},
  \bibinfo{journal}{\prl} \textbf{\bibinfo{volume}{92}},
  \bibinfo{pages}{018303} (\bibinfo{year}{2004}).

\bibitem[{\citenamefont{Chtchelkatchev
  et~al.}(2004)\citenamefont{Chtchelkatchev, Belzig, and
  Bruder}}]{chtchelkatchev04a}
\bibinfo{author}{\bibfnamefont{N.~M.} \bibnamefont{Chtchelkatchev}},
  \bibinfo{author}{\bibfnamefont{W.}~\bibnamefont{Belzig}}, \bibnamefont{and}
  \bibinfo{author}{\bibfnamefont{C.}~\bibnamefont{Bruder}},
  \bibinfo{journal}{\prb} \textbf{\bibinfo{volume}{70}},
  \bibinfo{pages}{193305} (\bibinfo{year}{2004}).

\bibitem[{\citenamefont{Flindt et~al.}(2005)\citenamefont{Flindt, Novotny, and
  Jauho}}]{jauho05}
\bibinfo{author}{\bibfnamefont{C.}~\bibnamefont{Flindt}},
  \bibinfo{author}{\bibfnamefont{T.}~\bibnamefont{Novotny}}, \bibnamefont{and}
  \bibinfo{author}{\bibfnamefont{A.-P.} \bibnamefont{Jauho}},
  \bibinfo{journal}{Europhys. Lett.} \textbf{\bibinfo{volume}{69}},
  \bibinfo{pages}{475} (\bibinfo{year}{2005}).

\bibitem[{\citenamefont{Usmani et~al.}(2006)\citenamefont{Usmani, Blanter, and
  Nazarov}}]{blanter06}
\bibinfo{author}{\bibfnamefont{O.}~\bibnamefont{Usmani}},
  \bibinfo{author}{\bibfnamefont{Y.~M.} \bibnamefont{Blanter}},
  \bibnamefont{and} \bibinfo{author}{\bibfnamefont{Y.~V.}
  \bibnamefont{Nazarov}}, \bibinfo{journal}{cond-mat/0603017}
  (\bibinfo{year}{2006}).

\bibitem[{\citenamefont{Haupt et~al.}(2006)\citenamefont{Haupt, Cavaliere,
  Fazio, and Sassetti}}]{fazio06}
\bibinfo{author}{\bibfnamefont{F.}~\bibnamefont{Haupt}},
  \bibinfo{author}{\bibfnamefont{F.}~\bibnamefont{Cavaliere}},
  \bibinfo{author}{\bibfnamefont{R.}~\bibnamefont{Fazio}}, \bibnamefont{and}
  \bibinfo{author}{\bibfnamefont{M.}~\bibnamefont{Sassetti}},
  \bibinfo{journal}{cond-mat/0607080}  (\bibinfo{year}{2006}).

\bibitem[{\citenamefont{Clerk}(2004)}]{clerk04}
\bibinfo{author}{\bibfnamefont{A.}~\bibnamefont{Clerk}},
  \bibinfo{journal}{\prb} \textbf{\bibinfo{volume}{70}},
  \bibinfo{pages}{245306} (\bibinfo{year}{2004}).

\bibitem[{\citenamefont{Cleland and Roukes}(1996)}]{roukes96}
\bibinfo{author}{\bibfnamefont{A.}~\bibnamefont{Cleland}} \bibnamefont{and}
  \bibinfo{author}{\bibfnamefont{M.}~\bibnamefont{Roukes}},
  \bibinfo{journal}{Appl. Phys. Lett.} \textbf{\bibinfo{volume}{69}},
  \bibinfo{pages}{2653} (\bibinfo{year}{1996}).

\bibitem[{\citenamefont{Knobel and Cleland}(2003)}]{knobel03a}
\bibinfo{author}{\bibfnamefont{R.~G.} \bibnamefont{Knobel}} \bibnamefont{and}
  \bibinfo{author}{\bibfnamefont{A.~N.} \bibnamefont{Cleland}},
  \bibinfo{journal}{Nature} \textbf{\bibinfo{volume}{424}},
  \bibinfo{pages}{291} (\bibinfo{year}{2003}).

\bibitem[{\citenamefont{Sapmaz et~al.}(2003)\citenamefont{Sapmaz, Blanter,
  Gurevich, and van~der Zant}}]{sapmaz03a}
\bibinfo{author}{\bibfnamefont{S.}~\bibnamefont{Sapmaz}},
  \bibinfo{author}{\bibfnamefont{Y.~M.} \bibnamefont{Blanter}},
  \bibinfo{author}{\bibfnamefont{L.}~\bibnamefont{Gurevich}}, \bibnamefont{and}
  \bibinfo{author}{\bibfnamefont{H.~S.~J.} \bibnamefont{van~der Zant}},
  \bibinfo{journal}{\prb} \textbf{\bibinfo{volume}{67}},
  \bibinfo{pages}{235414} (\bibinfo{year}{2003}).

\bibitem[{\citenamefont{Sazonova et~al.}(2004)\citenamefont{Sazonova, Yaish,
  Ustunel, Roundy, Arias, and McEuen}}]{mceuen}
\bibinfo{author}{\bibfnamefont{V.}~\bibnamefont{Sazonova}},
  \bibinfo{author}{\bibfnamefont{Y.}~\bibnamefont{Yaish}},
  \bibinfo{author}{\bibfnamefont{H.}~\bibnamefont{Ustunel}},
  \bibinfo{author}{\bibfnamefont{D.}~\bibnamefont{Roundy}},
  \bibinfo{author}{\bibfnamefont{T.~A.} \bibnamefont{Arias}}, \bibnamefont{and}
  \bibinfo{author}{\bibfnamefont{P.}~\bibnamefont{McEuen}},
  \bibinfo{journal}{Nature} \textbf{\bibinfo{volume}{431}},
  \bibinfo{pages}{284} (\bibinfo{year}{2004}).

\bibitem[{\citenamefont{Blencowe}(2005)}]{blencowe05a}
\bibinfo{author}{\bibfnamefont{M.~P.} \bibnamefont{Blencowe}},
  \bibinfo{journal}{Contemp. Phys.} \textbf{\bibinfo{volume}{46}},
  \bibinfo{pages}{249} (\bibinfo{year}{2005}).

\bibitem[{\citenamefont{Averin et~al.}(1991)\citenamefont{Averin, Korotkov, and
  Likharev}}]{averin91a}
\bibinfo{author}{\bibfnamefont{D.~V.} \bibnamefont{Averin}},
  \bibinfo{author}{\bibfnamefont{A.~N.} \bibnamefont{Korotkov}},
  \bibnamefont{and} \bibinfo{author}{\bibfnamefont{K.~K.}
  \bibnamefont{Likharev}}, \bibinfo{journal}{\prb}
  \textbf{\bibinfo{volume}{44}}, \bibinfo{pages}{6199} (\bibinfo{year}{1991}).

\bibitem[{\citenamefont{Beenakker}(1991)}]{beenakker91a}
\bibinfo{author}{\bibfnamefont{C.~W.~J.} \bibnamefont{Beenakker}},
  \bibinfo{journal}{\prb} \textbf{\bibinfo{volume}{44}}, \bibinfo{pages}{1646}
  (\bibinfo{year}{1991}).

\bibitem[{foo({\natexlab{a}})}]{footnote1}
\bibinfo{note}{These coefficients were called $\Gamma_L$, $\Gamma_R$ in
  Ref.~\onlinecite{armour04a}. Since they are {\it not} rates (e.g., they can
  become negative in the strong-coupling limit), we have chosen a different
  notation.}

\bibitem[{\citenamefont{Hershfield et~al.}(1993)\citenamefont{Hershfield,
  Davies, Hyldgaard, Stanton, and Wilkins}}]{hershfield93a}
\bibinfo{author}{\bibfnamefont{S.}~\bibnamefont{Hershfield}},
  \bibinfo{author}{\bibfnamefont{J.~H.} \bibnamefont{Davies}},
  \bibinfo{author}{\bibfnamefont{P.}~\bibnamefont{Hyldgaard}},
  \bibinfo{author}{\bibfnamefont{C.~J.} \bibnamefont{Stanton}},
  \bibnamefont{and} \bibinfo{author}{\bibfnamefont{J.~W.}
  \bibnamefont{Wilkins}}, \bibinfo{journal}{\prb}
  \textbf{\bibinfo{volume}{47}}, \bibinfo{pages}{1967} (\bibinfo{year}{1993}).

\bibitem[{\citenamefont{Korotkov}(1994)}]{korotkov94}
\bibinfo{author}{\bibfnamefont{A.}~\bibnamefont{Korotkov}},
  \bibinfo{journal}{\prb} \textbf{\bibinfo{volume}{49}}, \bibinfo{pages}{10381}
  (\bibinfo{year}{1994}).

\bibitem[{\citenamefont{Nazarov}(2003)}]{nazarov03a}
\bibinfo{author}{\bibfnamefont{Y.~V.} \bibnamefont{Nazarov}},
  \emph{\bibinfo{title}{Quantum Noise in Mesoscopic Physics}}
  (\bibinfo{publisher}{Kluwer}, \bibinfo{address}{Dordrecht},
  \bibinfo{year}{2003}).

\bibitem[{\citenamefont{Bagrets and Nazarov}(2003)}]{bagrets03a}
\bibinfo{author}{\bibfnamefont{D.~A.} \bibnamefont{Bagrets}} \bibnamefont{and}
  \bibinfo{author}{\bibfnamefont{Y.~V.} \bibnamefont{Nazarov}},
  \bibinfo{journal}{\prb} \textbf{\bibinfo{volume}{67}},
  \bibinfo{pages}{085316} (\bibinfo{year}{2003}).

\bibitem[{\citenamefont{Belzig}(2005)}]{belzig}
\bibinfo{author}{\bibfnamefont{W.}~\bibnamefont{Belzig}},
  \bibinfo{journal}{\prb} \textbf{\bibinfo{volume}{71}},
  \bibinfo{pages}{161301(R)} (\bibinfo{year}{2005}).

\bibitem[{foo({\natexlab{b}})}]{footnote2}
\bibinfo{note}{These equations agree with the system of equations shown in the
  Appendix of Ref. \onlinecite{armour04b}; the factor $k^2$ displayed there is
  a typo and should read $k$.}

\bibitem[{\citenamefont{Rodrigues and Armour}(2005)}]{rodrigues05a}
\bibinfo{author}{\bibfnamefont{D.~A.} \bibnamefont{Rodrigues}}
  \bibnamefont{and} \bibinfo{author}{\bibfnamefont{A.~D.}
  \bibnamefont{Armour}}, \bibinfo{journal}{\prb} \textbf{\bibinfo{volume}{72}},
  \bibinfo{pages}{085324} (\bibinfo{year}{2005}).

\bibitem[{\citenamefont{Blanter and B{\"u}ttiker}(2000)}]{blanter00a}
\bibinfo{author}{\bibfnamefont{Y.~M.} \bibnamefont{Blanter}} \bibnamefont{and}
  \bibinfo{author}{\bibfnamefont{M.}~\bibnamefont{B{\"u}ttiker}},
  \bibinfo{journal}{Phys. Rep.} \textbf{\bibinfo{volume}{336}},
  \bibinfo{pages}{1} (\bibinfo{year}{2000}).

\bibitem[{\citenamefont{Gustavsson et~al.}(2006)\citenamefont{Gustavsson,
  Leturcq, Simovic, Schleser, Ihn, Studerus, Ensslin, Driscoll, and
  Gossard}}]{gustavsson06a}
\bibinfo{author}{\bibfnamefont{S.}~\bibnamefont{Gustavsson}},
  \bibinfo{author}{\bibfnamefont{R.}~\bibnamefont{Leturcq}},
  \bibinfo{author}{\bibfnamefont{B.}~\bibnamefont{Simovic}},
  \bibinfo{author}{\bibfnamefont{R.}~\bibnamefont{Schleser}},
  \bibinfo{author}{\bibfnamefont{T.}~\bibnamefont{Ihn}},
  \bibinfo{author}{\bibfnamefont{P.}~\bibnamefont{Studerus}},
  \bibinfo{author}{\bibfnamefont{K.}~\bibnamefont{Ensslin}},
  \bibinfo{author}{\bibfnamefont{D.~C.} \bibnamefont{Driscoll}},
  \bibnamefont{and} \bibinfo{author}{\bibfnamefont{A.~C.}
  \bibnamefont{Gossard}}, \bibinfo{journal}{\prl}
  \textbf{\bibinfo{volume}{96}}, \bibinfo{pages}{076605}
  (\bibinfo{year}{2006}).

\bibitem[{\citenamefont{MacDonald}(1962)}]{macdonald62a}
\bibinfo{author}{\bibfnamefont{D.~K.~C.} \bibnamefont{MacDonald}},
  \emph{\bibinfo{title}{Noise and fluctuations: an introduction}}
  (\bibinfo{publisher}{Wiley}, \bibinfo{address}{New York},
  \bibinfo{year}{1962}).

\bibitem[{\citenamefont{Cottet et~al.}(2004{\natexlab{a}})\citenamefont{Cottet,
  Belzig, and Bruder}}]{cottet04:prl}
\bibinfo{author}{\bibfnamefont{A.}~\bibnamefont{Cottet}},
  \bibinfo{author}{\bibfnamefont{W.}~\bibnamefont{Belzig}}, \bibnamefont{and}
  \bibinfo{author}{\bibfnamefont{C.}~\bibnamefont{Bruder}},
  \bibinfo{journal}{\prl} \textbf{\bibinfo{volume}{92}},
  \bibinfo{pages}{206801} (\bibinfo{year}{2004}{\natexlab{a}}).

\bibitem[{\citenamefont{Cottet et~al.}(2004{\natexlab{b}})\citenamefont{Cottet,
  Belzig, and Bruder}}]{cottet04:prb}
\bibinfo{author}{\bibfnamefont{A.}~\bibnamefont{Cottet}},
  \bibinfo{author}{\bibfnamefont{W.}~\bibnamefont{Belzig}}, \bibnamefont{and}
  \bibinfo{author}{\bibfnamefont{C.}~\bibnamefont{Bruder}},
  \bibinfo{journal}{\prb} \textbf{\bibinfo{volume}{70}},
  \bibinfo{pages}{115315} (\bibinfo{year}{2004}{\natexlab{b}}).

\bibitem[{\citenamefont{Amman et~al.}(1989)\citenamefont{Amman, Mullen, and
  Ben-Jacob}}]{amman89a}
\bibinfo{author}{\bibfnamefont{M.}~\bibnamefont{Amman}},
  \bibinfo{author}{\bibfnamefont{K.}~\bibnamefont{Mullen}}, \bibnamefont{and}
  \bibinfo{author}{\bibfnamefont{E.}~\bibnamefont{Ben-Jacob}},
  \bibinfo{journal}{J. Appl. Phys.} \textbf{\bibinfo{volume}{65}},
  \bibinfo{pages}{339} (\bibinfo{year}{1989}).

\end{thebibliography}
\end{document}